\documentstyle[12pt,tighten,aps,amsfonts,bbm]{revtex}
\title{Generalized Quantum Mechanics and \\ 
       Nonlinear Gauge Transformations\thanks{
       to appear in ``Symmetry in Science IX'', B.~Gruber (Ed.),
       Plenum Pub., New York 1997.}}
\author{Peter Nattermann}
\address{Institut f\"ur Theoretische Physik\\
         Technische Universit\"at Clausthal\\
         Germany\\
         E-mail: {\tt aspn@pt.tu-clausthal.de}}
\preprint{ASI-TPA/4/97}
\let\mathscr\mathcal
\date{}
\def\Ref#1{(\ref{#1})}
\def\op#1{{\mathbf #1}}
\def\notion#1{{\bfseries\slshape #1}}
\def\dt{\partial_t}
\def\dx{\,d^3\!x}
\def\Ext{Ext}
\def\cz{{\mathbbm C}}
\def\gz{{\mathbbm Z}}
\def\rz{{\mathbbm R}}
\def\B{{\mathcal B}}
\def\C{{\mathcal C}}
\def\E{{\mathcal E}}
\def\G{{\mathcal G}}
\def\HS{{\mathscr H}}
\def\J{\vec{J}}

\def\MP{{\mathscr T}}
\def\M{{\mathscr M}}
\def\N{{\mathcal N}}
\def\O{{\mathcal O}}
\def\P{{\mathscr P}}
\def\S{{\mathcal S}}
\def\U{{\mathscr U}}
\def\X{{\mathfrak X}}

\def\x{\vec{x}}
\def\TC{{\cal T}_1^+}                 % trace-class op
\let\ds\displaystyle

\begin{document}
\maketitle
\begin{abstract}
Motivated by the problems of interpretation of a nonlinear evolution 
equation in quantum mechanics we discuss in this contribution 
the concept of \emph{nonlinear gauge transformations}, that has recently 
been introduced in joint work with 
{\scshape Doebner} and {\scshape Goldin}, in the framework of {\scshape
Mielnik}'s \emph{Generalized Quantum Mechanics}. 
Using these gauge transformations we construct linear quantum systems 
in a ``nonlinear disguise'', and a \emph{gauge generalization} of these  
(in analogy to the minimal coupling of orthodox quantum mechanics) leads 
to a unification of {\scshape Bialynicki-Birula--Mycielski} and 
{\scshape Doebner--Goldin} evolution equations for the quantum 
system.  
The notion of \emph{nonlinear observables} introduced by {\scshape L\"ucke} 
is finally discussed in the same framework.
\end{abstract}
\section{Introduction}\label{1}
There have been several approaches to a nonlinear modification of orthodox  
quantum mechanics and, in particular, of the {\scshape Schr\"odinger} 
equation. As the equations will appear in this paper we cite the nonlinear 
{\scshape Schr\"odinger} equations proposed by {\scshape Bialynicki-Birula} 
and {\scshape Mycielski} \cite{BiaMyc76},
\begin{equation}
\label{BBM}
  i\hbar\dt \psi_t = \left(-\frac{\hbar^2}{2m}\Delta + V\right) + \alpha_1
    \ln|\psi_t|^2\, \psi_t \,,
\end{equation}
and the one derived by {\scshape Doebner} and {\scshape Goldin} 
\cite{DoeGol92,DoeGol93b,DoeGol93a,DoeGol94} from the 
representation theory of the \emph{kinematical algebra} for a
single particle on $\rz^3$, 
\begin{displaymath}
  S(\rz^3) = \X (\rz^3) \oplus_{\mathcal L} C^\infty(\rz^3)\,. 
\end{displaymath}
Their family of equations depends on a real quantum number 
$D\in\rz$ obtained from the 
classification of unitary representations of $S(\rz^3)$ 
\cite{GoMeSh81b,AnDoTo83}, and five model parameter\footnote{
  $D'$ being a constant with the dimensions of $D$, to make $c_j$
  dimensionless} $c_j\in\rz$,
\begin{equation}
\label{DG}
  i\hbar\dt \psi_t = \left(-\frac{\hbar^2}{2m}\Delta + V\right)\psi_t +
  i\frac{\hbar D}{2} \frac{\Delta\rho_t}{\rho_t}\psi_t + \hbar D'
  \sum_{j=1}^5 c_j R_j[\psi_t]\,\psi_t
\end{equation}
where ($\rho_t = \psi_t\bar\psi_t$, $\J_t = {\mathrm Im}(\bar\psi_t
\vec{\nabla}\psi_t)$) 
\begin{equation}
\label{Rj}
\begin{array}{c}
  \ds  R_1[\psi] := \frac{\vec{\nabla}\cdot \J}{\rho}\,,\qquad
  R_2[\psi] := \frac{\Delta\rho}{\rho}\,,\qquad
  R_3[\psi] := \frac{\J^{\,2}}{\rho^2}\,, \\
  \ds  R_4[\psi] := \frac{\J\cdot\vec{\nabla}\rho}{\rho^2}\,,\qquad 
  R_5[\psi] := \frac{\vec{\nabla}\rho\cdot\vec{\nabla}\rho}{\rho^2}\,.
\end{array}
\end{equation}
For a further discussion of these equations we refer to \cite{Doebne97} in
these proceedings.

Along with {\scshape Weinberg}'s approach to a nonlinear modification 
of orthodox quantum mechanics \cite{Weinbe89a,Weinbe89b} 
there have been many discussions on the physical implications of a 
nonlinear quantum theory. In particular, {\scshape Gisin} 
\cite{Gisin90} and {\scshape Polchinski} \cite{Polchi91} pointed out 
that --- by an EPR-type experiment --- a nonlinear evolution equation 
would give us the opportunity of faster than light communications. 
Although this argument might not be persuasive in a
\emph{non-relativistic} theory, the structural implications of this
effect led to {\scshape Weinberg}'s sceptical attitude \cite{Weinbe92:book}. 

However, their argument is based on the notion of observables 
(as linear operators) and (mixed)
states (as density matrices) of orthodox, \emph{linear} quantum mechanics. 
The inconsistency of this notion for a nonlinear time evolution of
pure states becomes most evident for the time evolution of mixtures:  
Density matrices may be expanded in general \emph{non-uniquely} in terms of 
weights $ \lambda_j^{(\prime)}$ and eigenfunctions $ \psi_j^{(\prime)}$,
\begin{equation}
  \op{W} = \sum_{j} \lambda_j \vert \psi_j\rangle\langle \psi_j\vert
         = \sum_{j} \lambda_j^\prime \vert \psi_j^\prime\rangle\langle 
            \psi_j^\prime\vert \,.
\end{equation}
But evidently under a \emph{nonlinear} evolution of the pure states
$\psi_j$ and $\psi_j^\prime$ the two expansion evolve differently,
\begin{equation}
  \sum_{j} \lambda_j \vert \psi_j(t)\rangle\langle \psi_j(t)\vert
  \neq \sum_{j} \lambda_j^\prime \vert \psi_j^\prime(t)\rangle\langle 
            \psi_j^\prime(t)\vert \,,
\end{equation}
and there is no consistently defined time-evolved density matrix 
$\op{W}_t$!

Rather than interpreting this inconsistency as a source for
supraluminal communications we need a generalized concept of observables
and mixtures for nonlinear time evolutions. 

A framework for such a generalization is due to {\scshape Mielnik} 
\cite{Mielni74} and  
is based on the observation that {\sl ``what one observes in reality is a
distinguished role of the position measurement''}. Section 2 provides a 
brief review of {\scshape Mielnik}'s \emph{Generalized Quantum
Mechanics} along with two examples. 

Similar considerations on the distinguished role of positional measurements 
have recently led to the notion of \emph{nonlinear gauge 
transformation} in nonlinear quantum mechanics 
\cite{DoGoNa95,DoeGol96,DoGoNa96}, see also \cite{Goldin97} in these 
proceedings. 
In section 3 we merge these two ideas 
and introduce \emph{generalized gauge transformations} in generalized 
quantum mechanics.  

A particular, \emph{strictly local} class of generalized gauge 
transformations on $L^2(\rz^3,\dx)$ is identified as the class of 
nonlinear gauge 
transformations previously introduced \cite{DoGoNa95,DoeGol96,DoGoNa96}. 
They are used in section \ref{4} to obtain linear quantum systems in a 
\emph{nonlinear disguise}; their \emph{gauge generalization} (in analogy 
to minimal coupling in orthodox quantum mechanics) leads to a unification 
of {\scshape Bialynicki-Birula--Mycielski} \Ref{BBM} and {\scshape
Doebner--Goldin} equations \Ref{DG}.

Using these transformations {\scshape L\"ucke} \cite{Luecke95} 
proposed a generalized concept of \emph{nonlinear observables}
slightly different from {\scshape Mielnik}'s. The
concept and its relation to generalized quantum mechanics will be the
content of section \ref{5}.

\section{Generalized quantum mechanics}\label{2}
As indicated in the introduction we are looking for a quantum theory 
capable of nonlinear time evolutions of pure states. 
{\scshape Mielnik}'s concept for \emph{generalized quantum mechanics} 
provides a framework for such a theory. It has three basic ingredients.

The first ingredient has to be a ``model space'' for the pure
states. For a general setting we will take a topological manifold
$\MP$ as the \notion{manifold of pure states}. The topology of $\MP$ will
provide a notion as to which states are physically ``similar'', and thus
it will have to be determined by the physical interpretation of the
pure states. We should stress here, and this will be the case in the
examples below, that the points of $\MP$ are not necessarily in
one--to--one correspondence to the pure states.

The second ingredient is the set of time evolutions. 
In any experiment the experimentalist influences
the system by changing the external conditions. For a
non-relativistic (semi-classical) theory we might think of external
classical fields being applied, changing the evolution of the
system. We describe this dependence by elements \Ext\ of the
set $\C$ of all external conditions that are possibly applicable.
Thus in general the time evolutions are given by homeomorphisms of $ \MP$,
\begin{equation}
  T_{t,t^\prime,\Ext}:\, \MP \to \MP \,. 
\end{equation}
depending on the initial time $t$, the final time $t'$, 
and the external conditions  $\Ext\in\C$. 
Now the experimentalist will be free to change the external conditions
at different times, and the combination of these evolutions will have
to be considered as a possible evolution of the system as well. Thus
we call the smallest group containing \emph{all} such homeomorphisms,
closed in the topology of pointwise convergence\footnote{i.e.\ we
allow for infinitesimal, shock-like evolutions, see \cite{Mielni77}.}, 
the \notion{motion
group} $\M$ of the system.   

Finally, we need a notion of observables. As shown in the introduction
the set of observables will have to be invariant under time
evolutions, i.e.\ under the motion group $\M$, and we may generated it
closing a set of \notion{primitive observables} $\P$ under the action
of the motion group.
As mentioned by various
authors (see e.g.\ {\scshape Feynman} and {\scshape Hibbs}
\cite[p.~91]{FeyHib65:book} or {\scshape Mielnik} \cite{Mielni74})
the positional measurement are distinguished in a non-relativistic
quantum theory, and all other observables can be 
obtained from positional measurements. 
Adopting these observations we consider as
primitive observables the set $\P$ of positional observables 
\begin{equation}
  p_B : \MP \to [0,1]\,,
\end{equation}
labeled by {\scshape Borel}-sets $B \in\B(M)$ and interpreted as 
the probability of finding the system in the region $B\subset M$ of
physical space $M$,
i.e.\ for every $\phi\in\MP$, $p_{.}(\phi): \B(M)\to [0,1]$ is a
probability measure, i.e.\
\begin{equation}
  P_M = 1\,,\quad\mbox{and}\quad
  p_{B_1\cup B_2} = p_{B_1} + p_{B_2} \,, \quad\forall 
  B_1 \cap B_2 = \emptyset \,.
\end{equation} 

Using these notions we define a \notion{quantum system} to be a triple 
$(\MP,\M,\P)$ of a manifold of pure states $\MP$, a motion group 
$\M$, and a set of positional (primitive) observables $\P$.

For any such quantum system $(\MP,\M,\P)$ we are able to define
observables and mixed states as \emph{derived} concepts. 

Continuing the above argument any physical observable can be
obtained as a combination of time evolutions and a subsequent
positional measurement, i.e.\ 
the \notion{set of observables} is the completion\footnote{
  i.e.\ again we allow for infinitesimal combinations.} in the topology of 
pointwise convergence (p.c.) of the set of all combinations  
of time evolutions (under various external conditions) 
and positional (primitive) measurements\footnote{
  Here we deviate from {\scshape Mielnik} \cite{Mielni74} who took the
  \emph{linear span} for these combinations},
\begin{equation}
\label{O}
  \O := \overline{\left\{ p\circ T \left| p\in\P,
    T\in\M\right\}\right. }^{p.c.} \,.
\end{equation}

The set of mixed states of the quantum system $(\MP,\M,\P)$ depends on 
$\O$ in an obvious manner: Consider first statistical mixtures of pure 
states, i.e.\ probability measures $\pi$ on the pure state manifold $\MP$. 
For any observable $f\in \O$ the measure $\pi$ defines an expectation value 
\begin{equation}
  \pi(f):= \int_{\MP} f(\phi) d\pi(\phi)\,.
\end{equation}  
The first requirement on $\pi$ has to be that these expectation
values are finite for all observables, i.e.\ we consider the set of
finite probability measures  
\begin{equation}
  \Pi(\MP) := \left\{ \pi \mbox{ prob.~measure on }\MP \left| 
    \pi(f) < \infty\quad \forall 
    f\in\O\right.\right\}\,.
\end{equation}
Furthermore, we have to identify those probability measures 
that are indistinguishable by any kind of measurement, i.e.\ lead 
to the same expectation values for all observables. 
This establishes an equivalence relation among the 
finite probability measures,
\begin{equation}
    \pi_{1}\sim \pi_{2}  \quad\Leftrightarrow\quad  \pi_{1}(f) =  
    \pi_{2}(f)\quad\forall f\in\O\,, 
\end{equation}
and the \notion{set of mixed states} or \emph{statistical figure}
of the quantum system $(\MP,\M,\P)$ is the set of equivalence classes
\begin{equation}
  \S := \Pi(\MP) / \sim\,.
\end{equation}
The set $\S$ inherits a natural convex structure from 
$\Pi(\MP)$, and the \notion{pure states} are in one--to--one correspondence 
to the extremal points $\E$ of $\S$. 

Summing up, the concept provides a consistent description of
observables and mixed states for a quantum system with any sort (linear or 
nonlinear) of time evolutions. 

Before we turn to two examples, we note, that in \emph{classical}
statistical mechanics the equivalence relation would be
\emph{trivial}; the statistical figure there --- generated by probability 
measures on phase space --- is a (generalized)
simplex, i.e.\ every state has a unique decomposition in terms of pure
states (single point measure). 
Hence, a non-trivial equivalence relation reveals one characteristics 
of quantum mechanics.

\paragraph*{Example 1:}
Orthodox quantum mechanics can be described in terms of 
generalized quantum mechanics \cite{Mielni74}. For a single
particle in $\rz^3$ we consider a \emph{linear} quantum system 
$(\HS,U(\HS),\P)$. Here 
the \emph{manifold of pure states} is a separable {\scshape Hilbert}
space in a position representation,
\begin{equation}
\label{HS}
  \MP \equiv \HS \equiv L^2(\rz^3,\dx)\,.
\end{equation}
Thus for all non-zero elements $\psi$ of $\HS$ $|\psi(\x)|^2 / \|\psi\|^2$ is
a positional probability distribution, i.e.\ the 
\emph{position observables} are expectation values of 
characteristic functions $\chi_B$:
\begin{equation}
\label{pB}
    p_B[\psi] := \frac{\langle\psi\vert\chi_B\psi\rangle}{\|\psi\|^2} 
               = \frac{\int_B \bar\psi(\x) \psi(\x) \dx}{
                       \int_{\rz^3} \bar\psi(\x) \psi(\x) \dx}\,.
\end{equation}
The \emph{time evolutions} are generated by linear {\scshape Schr\"odinger}
equations 
\begin{equation}
\label{SE} 
  i\hbar\dt \psi_t = \left(-\frac{\hbar^2}{2m} \Delta +V\right)\psi_t\,,
\end{equation}
where the scalar potential\footnote{
  One may add a magnetic vector potential as
  well, but it turns out that the scalar potentials are already large
  enough to give the full unitary group as a motion group of the system}
$V$ represents a (sufficiently large) class of
external conditions $\C$.  
If we allow for arbitrary smooth potentials $V$ 
the \emph{motion group} is the unitary group on $\HS$ 
\cite{Mielni74,Waniew77}
\begin{equation}
\label{Mlin}
    \M = U(\HS)\,.
\end{equation}
As a consequence the \emph{set of observables} is given by the set of
orthogonal projectors on $\HS$,
\begin{equation}
\label{Olin}
  \O \simeq \textit{Proj }(\HS)\,,
\end{equation}
i.e.\ for all observables $o\in O$ there is a projector $\op{E}\in
\textit{Proj }(\HS)$ such that
\begin{equation}
\label{o}
  o[\psi] =
  \frac{\left\langle\psi\vert\op{E}\psi\right\rangle}{\|\psi\|^2}
  = \frac{\|\op{E}\psi\|^2}{\|\psi\|^2}\,.
\end{equation}
Thus generalized quantum mechanics naturally yields the lattice of
closed subspace of $\HS$, the \emph{quantum logic} of
\emph{propositions} \cite{Luecke96}.
Then due to {\scshape Gleason}'s theorem, the \emph{mixed states} are density 
matrices and the \emph{pure states} one-dimensional projectors (rays)
\begin{equation}
\label{Slin}
  \S \simeq \TC(\HS) \,,\qquad  \E \simeq P(\HS)\,,
\end{equation}

\paragraph*{Example 2:} 
A variation of the previous example of a single particle in $\rz^3$ is due to
{\scshape Haag} and {\scshape Bannier} \cite{HaaBan78}. They consider
a quantum system $(\HS,\M,\P)$, where $\HS$ and $\P$ are as in
\Ref{HS} and \Ref{pB}, respectively.
But for the \emph{time evolutions} they include a 
nonlinear term in the {\scshape Schr\"odinger} equation, 
\begin{equation} 
  i\hbar\dt \psi_t = \left(-\frac{\hbar^2}{2m} \Delta +V\right)\psi_t
   + \vec{A}\cdot \frac{\J_t}{\rho_t}\psi_t\,.
\end{equation}
Both, the scalar potential $V$ and the vector field $\vec{A}$
represent the external conditions $\C$ of the system.
Thus the \emph{motion group} still contains as a
sub-group the unitary group on $\HS$ (for $\vec{A}\equiv 0$), but is
effectively larger than in the purely linear case \Ref{Mlin},
\begin{equation}
  \M \supset U(\HS)\,.
\end{equation}
Consequently, the \emph{set of observables} is also larger 
than \Ref{Olin}, and contains non-quadratic forms as well, 
\begin{equation}
  \O \supset \textit{Proj } (\HS)\,.
\end{equation}
{\scshape Haag} and {\scshape Bannier} showed that the set of
observables for this quantum system is large enough to ``separate''
states, i.e.\ the equivalence relation of probability measures is 
trivial (on rays), and we get --- as opposed to \Ref{Slin} --- a
generalized simplex  
\begin{equation}
  \S = \Pi(\E)\,,\qquad \E = P(\HS)\,.
\end{equation}
Thus the quantum system $(\HS,\M,\P)$ has the \emph{classical}
feature of a unique decomposition of mixtures into pure components!

\section{Gauge equivalence}\label{3}
In orthodox quantum mechanics (on $\rz^3$) a \emph{gauge transformation} 
is a local unitary operator $\op{U}_\theta\in U_{loc}$
\begin{equation}
\label{Uloc}
  \left(\op{U}_\theta\psi_t\right)(\x) = e^{i\theta(\x,t)} 
  \psi_t(\x)\,, \quad \psi\in L^2(\rz^3,d^3 x)\,.
\end{equation}
We may describe $U_{loc}$ as the set of \emph{linear} transformations that 
leave the position probability invariant. 
These gauge transformations are employed for introducing 
classical electro-magnetic fields by replacing  $\dt\theta$ and
$\vec{\nabla}\theta$ obtained in the {\scshape Schr\"odinger} equation
through \Ref{Uloc} by more general scalar and vector potentials $V$
and $\vec{A}$, respectively, a process denoted as \notion{gauge 
generalization} in \cite{DoGoNa96}.

This notion of gauge transformations and gauge equivalence can 
naturally be adopted to the generalized setting of the previous
section: We define two quantum systems
$(\MP^{(1)},\M^{(1)},\P^{(1)})$ and  
$(\MP^{(2)},\M^{(2)},\P^{(2)})$ to be \notion{gauge equivalent}, 
iff
\begin{enumerate}
\item  $\P^{(1)}$ and $\P^{(2)}$ are positional observables over the 
    same space $M$;
\item  the external conditions $\C$ are the same,
\item  and there is a homeomorphism $N: \MP^{(1)}\to \MP^{(2)}$ such 
    that 
  \begin{displaymath}
  \begin{array}{rcll}
    \ds p^{(1)}_{B}& = &\ds  p^{(2)}_{B} \circ N\,,&\ds \quad \forall
    p^{(j)}\in   
    \P^{(j)}\,,\\
    \ds  T^{(1)}_{t,t',\Ext} & = &\ds N^{-1}\circ
    T^{(2)}_{t,t',\Ext}\circ N \,, & \ds \quad 
      \forall t,t'\in\rz,\,\Ext\in \C\,.
  \end{array}
  \end{displaymath}
\end{enumerate}  
Conditions 1 and 2 guarantee that the physical
interpretation of the observables is the same, whereas condition 3
establishes an isomorphism $N$ of the two quantum systems. 
We call $N$ a \notion{generalized gauge transformation} as it is a 
generalization of $\op{U}_\theta$ in orthodox quantum mechanics.    

\section{Gauge equivalence on $L^2(\rz^3,\dx)$ and gauge generalizations}
\label{4}
As we will see in this section the nonlinear gauge transformations 
introduced previously \cite{DoGoNa95,DoeGol96,DoGoNa96} are a special case 
of generalized gauge transformations. We can thus adopt the arguments of 
\cite{DoGoNa96} to obtain nonlinear quantum systems gauge equivalent to 
the linear quantum system, a gauge generalization of which yields 
a larger class of substantially nonlinear quantum systems.

We shall start with linear quantum mechanics, i.e.\
a quantum system $(\HS,\M,\P)$ with $\HS$, $\P$, and $\M$ given as in 
the example 1 of section \ref{2}, eqs.~\Ref{HS}--\Ref{Mlin}.
As explained in this example the set of pure states $\E$ of the
quantum system $(\HS,\M,\P)$ turns out to be the projective 
{\scshape Hilbert}-space $P(\HS)$, i.e.\ wave functions $\psi$ and
$c\psi,\,c\in\dot\cz$, in $\HS$ are identified. 
The generalized gauge transformation
\begin{equation}
  N\!: \HS \to \MP\,,
\end{equation}
that we are looking for, where $\MP$ is the manifold of pure states
of the nonlinear theory, will thus have to respect this structure 
in order to be consistent, so we assume
\begin{equation}
\label{proj}
  N[c\psi] = c'(c) N[\psi]\,,\qquad\forall c\in\dot{\cz}\,.
\end{equation}
Applying this relation twice to $c= c_1\,c_2$ leads to 
\emph{{\scshape Cauchy}'s power equation} on $\dot{\cz}$,
\begin{equation}
  c'(c_1c_2) = c'(c_1) c'(c_2)
\end{equation}
with the continuous solution
\begin{equation}
\label{cp}
  c'(c) = |c|^{\delta+i\gamma} \left(\frac{c}{\bar{c}}\right)^{
    \frac{\Lambda}{2}}\,,
\end{equation}
where $\delta,\gamma\in\rz$ and $\Lambda\in \gz$. To be more specific
we will have to fix the manifold of pure states $\MP$ of the
nonlinear theory we are aiming for. We take the simplest choice and
assume that the nonlinear theory is defined on the {\scshape Hilbert}
space,  
$$
  \MP = L^2(\rz^3,\dx)\,.
$$
Furthermore, we will restrict our attention to 
\emph{strictly local} gauge transformations, i.e.\
\begin{equation}
  N[\psi](\x) = n(\psi(\x),\x)\,,
\end{equation}
for a continuous function $n\!: \cz\times\rz^3\to\cz$. 
Identifying $n(1,\x) = \kappa(\x) \exp\big(i\theta(\x)\big)$ 
it follows immediately from \Ref{proj} and \Ref{cp} that $N$
depends on the parameters $\delta,\gamma,\Lambda$ and the two functions 
$\kappa:\rz^3 \to \rz$, $\theta:\rz^3 \to \rz$,
\begin{equation}
  N_{(\delta,\gamma,\Lambda;\kappa,\theta)}[\psi] = \kappa |\psi|^\delta
  \exp\Bigl\{i\left(
  \gamma \ln|\psi| + 
  \Lambda \arg\psi + \theta\right)\Bigr\}\,.
\end{equation}
For $N$ to be an invertible on $L^2(\rz^3,\dx)$,
\begin{equation}
  \delta = 1\,,\qquad \Lambda = \pm 1\,,
\end{equation}
and $\kappa$ has to be positive and bounded from both sides.
$N$ is even norm-preserving iff $\kappa\equiv 1$, and the strictly local 
norm-preserving generalized gauge transformations 
\begin{equation}
\label{NgLt}
  N_{(\gamma,\Lambda;\theta)}[\psi] = |\psi|
  \exp\Bigl\{i\left( \gamma \ln|\psi| + \Lambda \arg\psi \right)\Bigr\}\,.
\end{equation}
on $L^2(\rz^3,\dx)$ coincide with the nonlinear gauge 
transformations introduced\footnote{
  Actually, arbitrary, non-vanishing 
  $\Lambda$ have been considered there. Then $N$ will only be defined on some 
  subset of $L^2(\rz^3,\dx)$.}
in \cite{DoGoNa95,DoeGol96,DoGoNa96}.
In general, the parameter $\gamma$ and the function $\theta$
can be time dependent and the transformations \Ref{NgLt} are indeed 
continuous \cite{Luecke95b}, 
i.e.\ and thus $N_{(\gamma,\Lambda;\theta)}$ is a homeomorphism of 
$\MP=L^2(\rz^3,\dx)$.

The set of these generalized gauge transformations forms a
group $\G$ under compositions; it is actually a semi-direct product of
the groups of \emph{pure} ($\theta\equiv 0$) nonlinear gauge
transformations $\N$ and the local unitary transformations $U_{loc}$,
\begin{equation}
  \G = \N \otimes_{s} U_{loc}\,,
\end{equation}
the semi-direct group structure being given by 
\begin{equation}
  \op{U}_{\Lambda\theta} \circ N_{(\delta,\gamma,\Lambda,0)} =
  N_{(\delta,\gamma,\Lambda,0)} \circ \op{U}_{\theta}\,.
\end{equation}
In view of this group structure and the fact that $\Lambda=-1$
corresponds to complex conjugation, it suffices for our purpose 
to use purely nonlinear gauge 
transformations 
\begin{equation}
  N_\gamma (\psi_t) = \psi_t \exp\left(i\gamma_t\ln|\psi_t|\right)\,.
\end{equation}

Starting with linear {\scshape Schr\"odinger} equations \Ref{SE} 
we obtain for the transformed wave function $
\psi^\prime_t=N_\gamma[\psi]$ evolution equations of 
{ \scshape Doebner--Goldin} and {\scshape Bialynicki-Birula--Mycielski}
type: 
\begin{equation}
\label{lNSE}
\begin{array}{rcl}
\ds
  i\hbar \dt \psi_t^\prime &=&\ds \left(-\frac{\hbar^2}{2m} \Delta
+V\right)\psi_t^\prime 
 - \frac{1}{2}\dot\gamma_t \ln |\psi_t^\prime|^2\, \psi_t^\prime\\
    &&\ds  
     + \frac{\hbar^2 \gamma_t}{4m} \left(iR_2[\psi] 
      +2R_1[\psi_t^\prime]-2 R_4[\psi_t^\prime]\right) \psi_t^\prime
     - \frac{\hbar^2 \gamma_t^2}{8m}
       \left(2R_2[\psi_t^\prime]-R_5[\psi_t^\prime]\right)\psi_t^\prime 
\end{array}
\end{equation}
with $R_j[\psi]$ as in \Ref{Rj}. 

Note, that these equations are formal generators of the nonlinear time
evolution \cite{Luecke95,Natter94} 
\begin{equation}
  \U_t = N_\gamma \circ \op{U}(t) \circ N_{-\gamma}\,,
\end{equation} 
where $\op{U}(t)$ denotes the unitary time evolution of the
linear {\scshape Schr\"odinger} equation \Ref{SE}.
 
\emph{Gauge generalization} \cite{DoGoNa96} now involves breaking the constraints
among the coefficients of the various terms in \Ref{lNSE} a generalization of the 
parameters of the equation \Ref{lNSE}: the coefficient of
$iR_2[\psi]+R_1[\psi]-R_4[\psi]$ depends linearly on 
$\gamma_t$, whereas that of $R_2[\psi]-2R_5[\psi]$ is quadratic in
$\gamma_t$, and that of $\ln |\psi|^2$ is proportional to the derivative
of $\gamma_t$. Thus as in a first step we could choose
these three coefficients independently. This leads to a
family of equations 
\begin{equation}
\label{R1}
\begin{array}{rcl}
\ds  i \frac{\dt \psi}{\psi} -\mu_0 V &=&\ds
 \nu_1 \left(iR_1[\psi]+\frac{1}{2}R_2[\psi] -R_3[\psi] -\frac{1}{4}R_5[\psi]
    \right) \\
&&\ds+ {\mu_1}_t \left( iR_2[\psi]+R_1[\psi]-R_4[\psi]\right) 
    +  {\mu_2}_t \left(R_2[\psi]-2R_5[\psi]\right) + {\alpha_1}_t \ln
    |\psi_t|^2
\end{array} 
\end{equation}
that is invariant under gauge transformations $N_\gamma$. In \Ref{R1}
we have used the expansion 
\begin{equation}
    \frac{\Delta\psi}{\psi} = iR_1[\psi]+\frac{1}{2}R_2[\psi] -R_3[\psi]
    -\frac{1}{4}R_5[\psi] \,,
\end{equation}
and introduced new parameters $\nu_j$, $\mu_k$, and $\alpha_1$ in an
obvious way. 

Further steps of gauge generalization are possible.   
For instance, if we distinguish real and imaginary parts in
\Ref{R1} and close the resulting family of equations with respect to
the gauge group $\N$ we obtain  
a unified (and generalized) 
{\scshape Bialynicki-Birula--Mycielski--Doebner--Goldin} 
equation 
\begin{equation}
\label{BBMDG}
  i\frac{\dt \psi_t}{\psi_t} - \mu_0 V = i \sum_{j=1}^2 {\nu_j}_t R_j[\psi_t] 
  + \sum_{k=1}^5 {\mu_k}_t R_k[\psi_t] + {\alpha_1}_t \ln |\psi_t|^2
\end{equation}
with real time-dependent parameters.

By construction --- as in the linear theory --- 
the generalized gauge transformations
are still automorphisms of the set of nonlinear {\scshape
  Schr\"odinger}  equations
obtained.

\section{Nonlinear Observables}\label{5}
Having constructed the time evolutions of nonlinear quantum systems,
we would like to know the structure of the \emph{set of observables}
of these quantum systems.

To gain some insight we utilize the concept of generalized gauge
transformations and consider again a quantum system $(\HS,\M,\P)$ gauge
equivalent to a linear quantum system $(\HS,U(\HS),\P)$, i.e.\ $\HS$
and $\P$ are as in \Ref{HS} and \Ref{pB}, respectively, and the motion
group is 
\begin{equation}
 \M = \left.\left\{ N\circ \op{U} \circ N^{-1} \right| \op{U} \in
   U(\HS)\right\}\,, 
\end{equation}
where $N$ is a (not necessarily strictly local)
generalized gauge transformation on $\HS=L^2(\rz^3,\dx)$. 
However, we shall assume as in the previous section that $N$ is 
\emph{norm-preserving} on $\HS=L^2(\rz^3,\dx)$. 

According to \Ref{o} of example 1 the observables of
the linear quantum system $(\HS,U(\HS),\P)$ are associated to
projection operators $\op{E}$. For the nonlinear quantum system
$(\HS,\M,\P)$ the set of observables is then  
\begin{equation}
  \O_{nl} \simeq \left\{\hat{E}\!:\HS\to\HS\left|  
  \hat{E} := N\circ \op{E} \circ
  N^{-1},\, \op{E}\in\textit{Proj }(\HS)\right\}\right. \,. 
\end{equation}
The maps are 
idempotent, $\hat{E}\circ \hat{E} = \hat{E}$ and to every $\hat{E}$ there 
is a unique $\neg\hat{E}\in\O_{nl}$ such that $\neg\hat{E}\circ\hat{E} = 
\hat{E}\circ\neg\hat{E} = 0$. Thus $\hat{E}$ is a 
\notion{generalized projection} onto a nonlinear 
submanifold of $L^2(\rz^3,\dx)$. Furthermore, $\O_{nl}$ inherits the
logical structure of $\textit{Proj }(\HS)$, it is a 
nonlinear realization of a \emph{quantum logic}, and the motion group
is contained in its automorphism group. 

The expectation values of $\op{E}$ w.r.t.\ $\psi$ and $\hat{E}$
w.r.t.\ $N[\psi]$ are the same,
\begin{equation}
  \frac{\left\langle\psi\vert\op{E}\psi\right\rangle}{\|\psi\|^2}
  = \frac{\|\op{E}\psi\|^2}{\|\psi\|^2} = \frac{\|(\op{E}\circ
  N^{-1}\circ N)[\psi]\|^2}{\|\psi\|^2} =
  \frac{\|\hat{E}\big[N[\psi]\big]\|^2}{\|N[\psi]\|^2}\,.  
\end{equation}  
The standard definition of observables in linear
quantum theory, however, goes one step further and associates to each
observable $A$ a self-adjoint operator, or equivalently (by the
spectral theorem) a projection-valued measure 
\begin{equation}
  \op{E}^A\!: \B(\rz) \to \textit{Proj }(\HS)
\end{equation}
that determines the distribution of the values of the observable $A$
on the real line $\rz$. 
With the above definition of generalized projections the 
nonlinear analogue to these projection valued measures is a 
\notion{generalized projection-valued measure} \cite{Luecke95}, 
\begin{equation} 
  \hat{E}^A(.) := N \circ \op{E}^{A}(.) \circ N^{-1}\!: 
  \B(\rz) \to \O_{nl} \,.
\end{equation}
It inherits the following properties from the linear case: 
\begin{enumerate}
\item[(i)] %Statistical interpretation:
  \begin{equation}
    \mu_\psi\!: B \mapsto  \mu_\psi(B) :=
  \frac{\|\hat{E}^A(B)[\psi]\|^2}{\|\psi\|^2} 
  \end{equation}
  for all $B\in\B(\rz)$, $\psi\in \dot\HS$
  defines a probability measure on $\rz$;
\item[(ii)] 
  \begin{equation}
    \hat{E}^A(B_1) \circ \hat{E}^A(B_2) = \hat{E}^A(B_1\cap B_2)
  \end{equation}
  for all $B_1,B_2\in\B(\rz)$;
\item[(iii)] %ideal measurements 
  \begin{equation}
    \mu_\psi(B) = 1 \quad \Rightarrow \quad E^A(B)\psi =\psi
  \end{equation}
\end{enumerate}
If $E^A(B)$ may be interpreted as the probability of measuring 
a value of a physical observable in $B$, we may call the 
generalized projection-valued measure a \notion{nonlinear observable}. 

For linearizable quantum systems this generalized notion establishes
the full equivalence of the two descriptions \cite{Luecke95}:
\begin{center}
\begin{tabular}{c|c|c}
                   & nonlinear theory      & linear theory \\
\hline
  wave functions   & $\psi^\prime= N[\psi]$ & $ \psi$ \\
  time evolution   & $ \U_t = N\circ \op{U}_t\circ N^{-1}$
  & $ \op{U}_t = \exp(-\frac{i}{\hbar}\op{H}t)$ \\
  observables      & $\left\{{E^A}^\prime = N\circ \op{E}^A\circ
  N^{-1}\right\}$ & $\{\op{E}^A\},\, \op{A}\in L_{sa}(\HS)$\\
  position         & $ \{ \chi_B\}$ & $ \{ \chi_B\}$   
\end{tabular}
\end{center}
 
\section{Final Remarks}
The purpose of this contribution has been to merge {\scshape Mielnik}'s 
concept of generalized quantum mechanics with the concept of nonlinear 
gauge transformations. Having introduced generalized gauge transformations 
for the quantum systems of generalized quantum mechanics, we were able to 
identify nonlinear gauge transformation recently introduced in the 
context of {\scshape Doebner--Goldin} equations as strictly local, 
norm-preserving generalized gauge transformations on $L^2(\rz^3,\dx)$. 
On the one hand these nonlinear gauge transformations were 
subsequently employed to obtain a unified class 
of {\scshape Bialynicki-Birula--Mycielski} and {\scshape Doebner--Goldin} 
evolution equations. 

On the other hand generalized norm-preserving gauge transformations were used 
to construct a nonlinear (linearizable) realization of a quantum logic on 
$L^2(\rz^3,\dx)$, and led to the definition of generalized 
projection-valued measures and nonlinear observables. 
These may serve as candidates for observables in 
truly nonlinear quantum theories. Following the arguments of section
\ref{2} the set of all these nonlinear observables will have to be
invariant under \emph{all} time evolutions admissible. 
However, for the formal evolution equations obtained in 
section \ref{4} via gauge generalization 
these time evolutions are still an open problem. So far, the 
{\scshape Cauchy} problem for the {\scshape Doebner--Goldin} sub-family 
of \Ref{BBMDG} for certain time-independent coefficients can be solved 
only on a certain dense subset of non-vanishing wave functions in 
$L^2(\rz^3,\dx)$ \cite{Teisma97}. But then there cannot be any 
ideal position measurements in this theory as such a measurement would result
in a {\scshape Cauchy} initial wave function outside the domain of definition 
for a further time evolution.

Another motivation for the present article was the prediction of
supraluminal communications in nonlinear quantum theory. The review of
{\scshape Mielnik}'s generalized quantum mechanics in section \ref{2}
showed that the formal obstacle that led to this conclusion is due to
an improper definition of observables and mixtures. Furthermore, the
equivalent \emph{nonlinear} description of linear quantum mechanics of
section \ref{5} shows that there are (seemingly) nonlinear time
evolutions with no supraluminal communications at all! 

For truly nonlinear time evolutions, however, the question whether
supraluminal communications are possible remains unsolved. 
The crucial
point in order to answer this question will be the description of two
(or many) particle systems and, in particular, of correlations
in such a nonlinear theory\footnote{
  A hint in this direction has been made in
  \cite{HaaBan78} where the authors refrain from calling their theory
  classical because of problems of interpretation in a many particle
  theory.}. 

The formalism of gauge generalizations employed in section \ref{4} can
--- at least formally --- be extended to more general cases. For
instance, we could follow \cite{Mielni74} and consider as a manifold
of pure states of the nonlinear system the set of $k$-integrable 
complex functions $(k>0)$,
\begin{equation}
  \MP^{(k)} := \left\{ \psi:\rz^3\to\cz\left| n^{(k)} [\psi] := 
  \int_{\rz^3} |\psi(\x)|^k \dx < \infty \right\}\right. \,,
\end{equation}
where $|\psi(\x)|^k/n^{(k)}[\psi]$ is taken as a position-probability
density. The resulting strictly local gauge transformations and the
formal nonlinear {\scshape Schr\"odinger} equations will be treated
elsewhere \cite{Natter97b,DoGoNa97}.
 
\subsection*{Acknowledgments}
I am grateful to the organizers of ``Symmetry in Science IX'' in
Bregenz for their invitation, support, and hospitality.
Some of the results presented here have been obtained in collaboration 
with {\scshape H.-D.~Doebner} and {\scshape G.A.~Goldin}; thanks
are also due to them as well as {\scshape W.~L\"ucke} for valuable 
discussions on the subject.

\end{document}